# CuLD: Current-Limiting Differential Reading Circuit for Current-Based Compute-in-Memory

Seiji Uenohara*, Satoshi Awamura, and Norio Hattori

Nuvoton Technology Corporation Japan, Advanced Development Section, Technology Innovation Department
1 Katari-yakemachi, Nagaokakyo City, Kyoto 617-8520, Japan, *uenohara.seiji@nuvoton.com, *s.uenohara@gmail.com

**Abstract**
This paper proposes a circuit configuration that addresses the issue of deviation in the multiply-accumulate (MAC) results when numerous word lines are simultaneously opened in current-based compute-in-memory (CiM) circuits. The proposed circuit solves this problem by automatically shrinking the product value according to the degree of parallelism. This circuit configuration is effective for circuit methods that calculate MAC through time integration of charge.

**Introduction**

Non-volatile resistive memory (NVM) is attracting attention as a device that maintains neural network weights during power-off in AI edge devices and perform highly efficient multiply-accumulate (MAC) operations in the current domain [1-6]. In neural networks using NVM, weights are expressed as resistance values, and MAC operations are performed by passing current through resistive devices [7,8] (see Fig. 1). The MAC value, represented by the capacitor potential of the CR circuit, has low linearity due to its exponential change. Improving the linearity is one of the challenges in current-based CiM. Additionally, when the row parallelism $N$ is large, the parallel resistance of the CR circuit decreases, causing the capacitive charge to discharge rapidly, making it difficult to extract the MAC operation results as an analog quantity. This is also a challenge. This paper proposes a Current Limiting Differential Reading Circuit (CuLD) to address these challenges.

**Current Limiting Differential Reading Circuit (CuLD)**

The principal circuit of CuLD is shown as Fig.2(a). In this circuit, a weight with a sign consists of four 1T-1R memory cells. To read the MAC results as an analog quantity with large $N$, CuLD has the following features: 1) the total current on each -bit line is limited to $I_{bias}$, 2) word lines are driven by a pulse-width modulation (PWM) signal $WL_i$ and $WLB_i$, which is the inverted $WL_i$. Here, $i$ is the word line index. (1) is a constraint to ensure that the result of the MAC is represented within a certain range regardless of $N$, and (2) is a driving method necessitated by the constraint in (1). The result of the multiply-accumulate (MAC) operation is expressed by the voltage difference between the capacitors on the positive/negative side. The voltage difference across the capacitors after the completion of the PWM signal input is expressed as follows:

$$V_{x,p} - V_{x,n} = \frac{1}{C}\sum_{i=1}^{N}(2X_i - X_{max})(I_{p,i} - I_{n,i}),$$

where $X_i$ and $X_{max}$ are the pulse width of the PWM signal driving the word-line and the maximum pulse-width, respectively. This equation shows that the sign of $X_i$ is positive when $X_i > X_{max}/2$, and negative when $X_i < X_{max}/2$. To achieve ideal multiply-accumulate operations with CuLD, the resistance values of the diagonal resistors among the four memory cells must match, and the parallel resistance of adjacent resistors must all be the same. When this condition is satisfied, $I_{p,i}=I_{bias}R_{n,i}/N(R_{p,i}+R_{n,i})$ and $I_{n,i}=I_{bias}R_{p,i}/N(R_{p,i}+R_{n,i})$ (see Fig. 3 for the equivalent circuit, excluding current mirror circuits for sensing of current values on bit lines, from Fig. 2(a)). Here, $I_{p,i}$ and $I_{n,i}$ are the currents through resistors $R_{p,i}$ and $R_{n,i}$, respectively. Since $I_{p,i}$ and $I_{n,i}$ are scaled by $1/N$, theoretically, the result of MAC is represented within a constant voltage range regardless of $N$.

Next, the necessity of the word line driving method using $WL_i/WLB_i$ will be explained using the circuit in Fig. 4, which excludes current mirrors for sensing. Without $WLB_i$, the same total current flows through BLP/BLN during both Period A and Period B, and the current changes due to the switching of the PWM signal are not reflected in the MAC. On the other hand, with $WLB_i$, the current path is inverted (positive/negative) during the switching of the PWM signal, reflecting the changes in the signal in the current (see Table I).

**Circuit Simulation Results**

Fig. 5 shows voltage waveforms of the capacitors during the MAC operation for a conventional circuit (see Fig. 2(b)) and proposed circuit when $N$ is varied. When $N$ was varied, odd-numbered word lines were input with the same signal as $WL_1$, and even-numbered word lines were input with the same signal as $WL_2$. In the conventional circuit, when $N$=32, the difference in capacitor potential between the positive and negative sides is maintained at 100ns, but when $N$=1,024, the potential difference is almost zero. On the other hand, in the proposed circuit, the potential difference is maintained even when $N$=1,024. Additionally, the linearity of the change in capacitor potential is better than that of the conventional circuit.

Fig. 6 shows the change in the difference in capacitor potential when $N$ and the resistance values are varied under the same conditions as in Fig. 5. In the conventional circuit, the difference in capacitor potential rapidly decreases as $N$ increases, becoming zero at $N$=128. In contrast, the change in the difference in capacitor potential in the proposed circuit is more gradual than in the conventional circuit.

Fig. 7 shows the difference in capacitor potential when the pulse width $X_1$ of the input PWM and $N$ are varied at $X_{max}$=100ns. As $N$ increases, the slope becomes more gradual while maintaining linearity. The reason for the smaller slope is that increasing $N$ makes the influence of the current source's output resistance more apparent.

To observe the influence of the output resistance of the current source, we focus on the difference $I_{diff}$ between the bit line currents $I_p$ and $I_n$ when $I_{bias}$ is varied. The voltage dynamic range of the MAC in CuLD is determined by the difference in bit line currents on the positive and negative sides, the capacitance value, and the charge time. However, since the capacitance value and charge time are merely coefficients, we evaluate $I_{diff}$ when driving the word line at $X_{max}$ (see Fig. 8).

Fig. 9 shows the change in $I_{diff}$ when $N$ and $I_{bias}$ are varied. This result indicates that a smaller output resistance of the current source expands the dynamic range.

**Conclusion**

Table II summarizes the specifications of representative AI accelerators using NVM. Among the circuits in this table that can simultaneously activate hundreds of word lines (item (5)), CuLD is the only one that can represent input vectors using PWM signals, which are highly compatible with digital circuits. Additionally, while CuLD has the disadvantage of representing one weight with four cells, the number of effective weights (item (7)) is high, and the feature of automatically scaling the products value (item (8)) is unique. The proposed circuit can improve computational parallelism without limiting the device, as long as it is a current-based CiM circuit.

**Acknowledgements**
The paper is based on results obtained from project JPNP23015, subsidized by the New Energy and Industrial Technology Development Organization (NEDO).

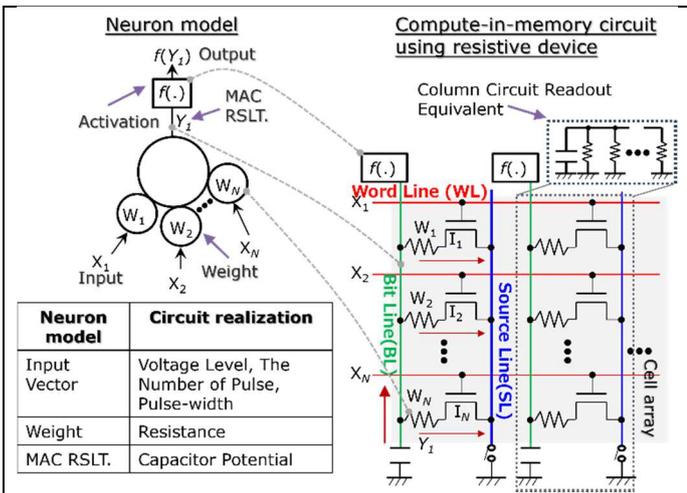

Fig. 1. Implementation of a neuron model using a CiM circuit with resistive memory. The weighted current driven by the selected WL flows through the bit line. When multiple WLs are selected simultaneously, the currents on the BL are summed, enabling MAC operations.

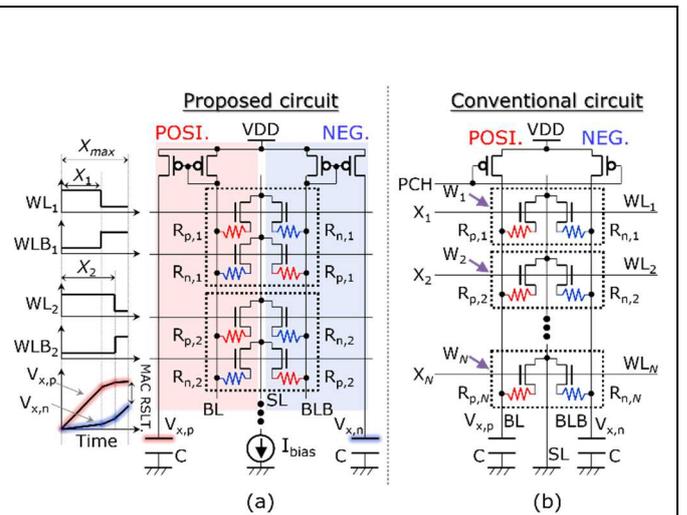

Fig. 2. (a) principle circuit of CuLD. The word lines are driven by positive/negative logic PWM signals. The result of the MAC operation is represented by the difference between the two capacitor voltages. (b) conventional circuit for comparison.

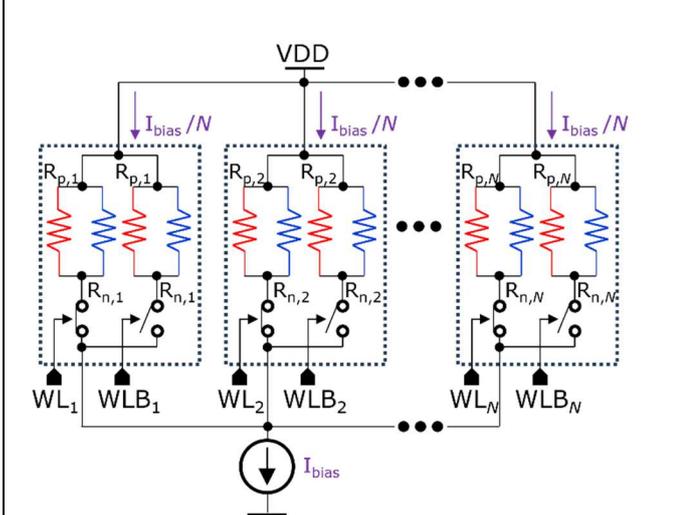

Fig. 3. CuLD's equivalent circuit without current mirror circuit for sensing. When the combined resistance values of the dashed resistors are all the same, the current flowing into the dashed block becomes $I_{bias}/N$.

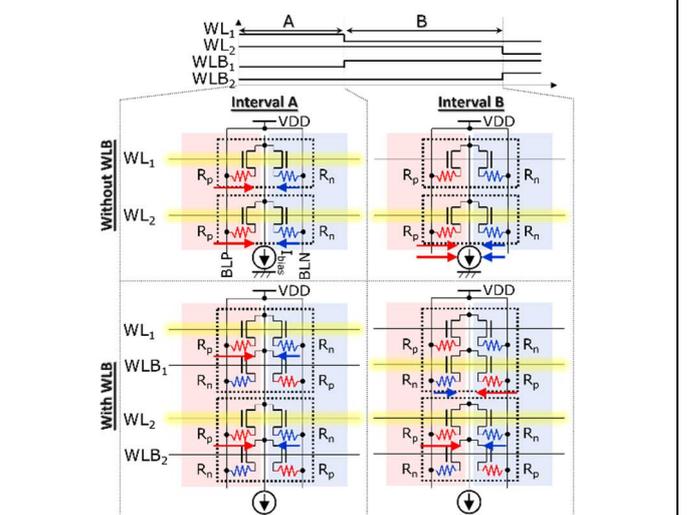

Fig. 4. The difference in the current flowing into BLP/BLN with and without $WLB_i$. This is an example for the case where $R_p < R_n$. Since the total current is determined by $I_{bias}$, the same current flows through the bit line in Period A/B without $WLB_i$.

Table I. Summary of the total current on bit lines BLP/BLN with/without $WLB_i$.

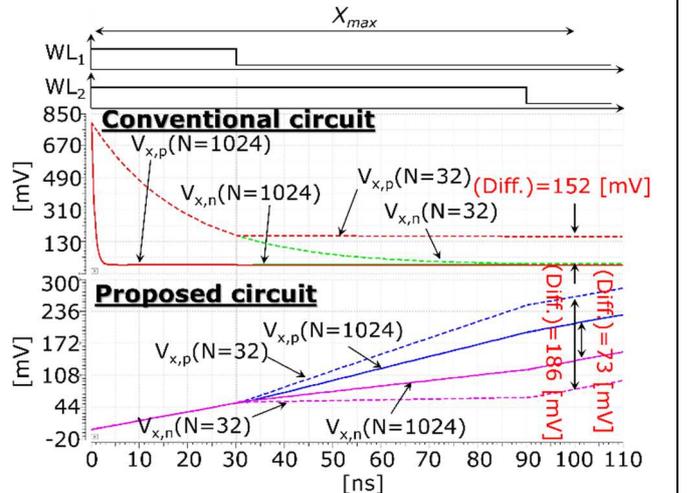

Fig. 5. Input PWM signal to the circuits in Figs. 1 and 2, and the time waveform of the capacitor potential. Circuit simulations were performed using the following parameters: VDD=0.8 V, temp. =25 °C, $R_{p,0}$=10 MΩ, $R_{n,0}$=100 kΩ, $R_{p,1}$=100 kΩ, $R_{n,1}$=10 MΩ, $I_{bias}$=10 μA, and C=3 pF.

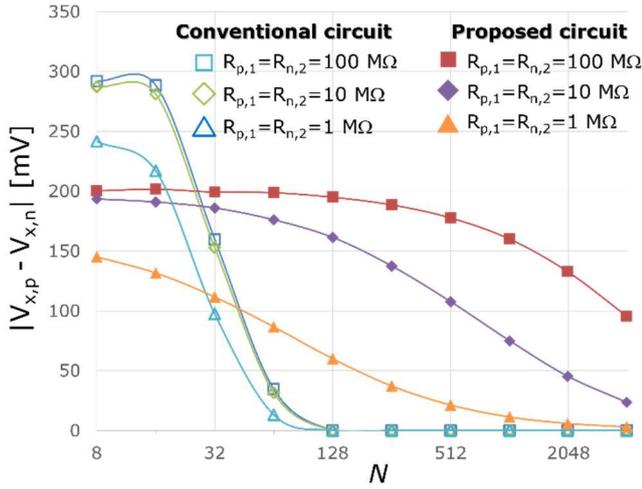

Fig. 6. Capacitor potential difference at 100ns when $N$ is varied. Circuit simulations were performed using the following parameters: VDD=0.8 V, temp. =25°C, $R_{p,1}=R_{n,0}$=100 kΩ, $I_{bias}$=10 μA, and C=3 pF.

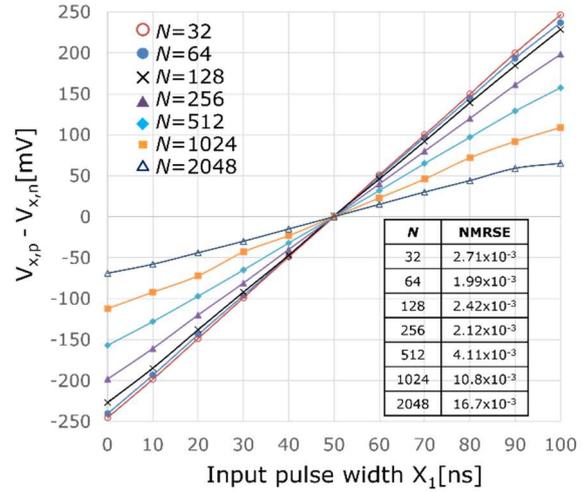

Fig. 7. Capacitor potential difference for each input value $X_0$ when $N$ is varied. Circuit simulations were performed using the following parameters: VDD=0.8 V, temp. =25°C, $R_{p,0}$=1 MΩ, $R_{n,0}$=10 MΩ, $I_{bias}$=10 μA, and C=3 pF. The vertical axis corresponds to the MAC operation result.

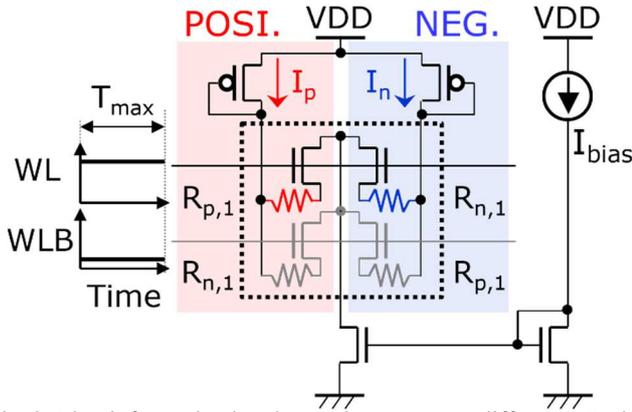

Fig. 8. Circuit for evaluating the maximum current difference. In this evaluation, only WL is turned ON with the maximum pulse width $T_{max}$. When $R_{p,0} < R_{n,0}$, $I_p$ is at its maximum and $I_n$ is at its minimum. The parameters were set as follows: $R_{p,0}$=1 MΩ, $R_{n,0}$=10 MΩ, $I_{bias}$=10 μA, and C=3 pF.

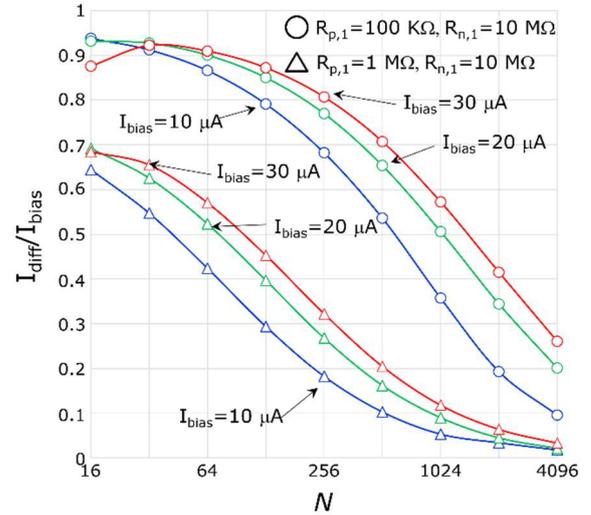

Fig. 9. Change in $I_{diff}$ when $I_{bias}$ is varied. Note that the vertical axis is scaled by $1/I_{bias}$. These results show that a larger $I_{bias}$ can achieve a greater $I_{diff}$ at a larger $N$. Additionally, when the resistance values of the high-resistance devices are the same, a smaller resistance value

TABLE II. Comparison of NVM Accelerators and CuLD

|  | [9] | [10] | [11] | [12] | CuLD (This work) |
|---|---|---|---|---|---|
| (1) Input Vector Representation | Analog voltage | 1bit pulse | Analog voltage | 1bit pulse | PWM |
| (2) Weight Storage | PCM | ReRAM | ReRAM | MRAM | ReRAM*1 |
| (3) Memory Cell Structure | 4T2R | 1T1R | 1T1R | 1T1R | 1T1R |
| (4) Weight Realization | 2 cells | 2 cells | 2 cells | 2 cells | 4 cells |
| (5) # of activated WLs | 256 | 1,568 | 256 | 256 | 1,024 or higher |
| (6) # of WLs per Weight | 1 | 2 | 1 | 1 | 2 |
| (7) # of Effective Inputs*2 | 256 | 784 | 256 | 256 | 512 or higher |
| (8) 1/$N$ Auto Scaling with Current Limit | NO | NO | NO | NO | YES |

*1 Our approach is device-agnostic as long as coefficients can be read out as current values.
*2 This value was calculated by (5)/(6).